\pdfoutput=1

\documentclass[aps,pra,10pt,onecolumn,superscriptaddress,nofootinbib]{revtex4}
\usepackage{amsmath}
\usepackage{latexsym}
\usepackage{amssymb}
\usepackage{graphics,epstopdf}

\begin{document}

\title{Lower bound of quantum uncertainty from extractable classical information}

\author{T. Pramanik}
\thanks{Tanumoy.Pramanik@telecom-paristech.fr}
\affiliation{LTCI, T´el´ecom ParisTech, 23 avenue dItalie, 75214 Paris CEDEX 13, France}

\author{S. Mal}
\thanks{shiladitya@bose.res.in} 
\affiliation{S. N. Bose National Centre for Basic Sciences, Salt Lake, Kolkata 700 098, India}

\author{A. S. Majumdar}
\thanks{archan@bose.res.in}
\affiliation{S. N. Bose National Centre for Basic Sciences, Salt Lake, Kolkata 700 098, India}

\date{\today}

\begin{abstract}

The sum of entropic uncertainties for the measurement of two 
non-commuting observables is not always reduced by the amount of 
entanglement (quantum memory) between two parties, and in certain cases may
be impacted by quantum correlations beyond entanglement (discord). An optimal 
lower bound of entropic uncertainty in the presence of any correlations may
be determined by fine-graining. Here we express the uncertainty relation in a new form
where the maximum possible reduction of uncertainty is shown to be given by the 
extractable classical information. We show that the lower bound of uncertainty
matches with that using fine-graining for several examples of two-qubit 
pure and mixed entangled states, and also separable states with 
non-vanishing discord.
 Using our uncertainty relation we further show that even in the 
absence of any quantum correlations between the two parties, the sum of 
uncertainties may be reduced with the help of classical correlations. \\\\


\end{abstract}

%\pacs{03.67.-a, 03.67.Mn}

\maketitle

\section{Introduction}

A fundamental difference from classical theory is that quantum theory limits 
the precision of the measurement outcomes for the measurement of two 
non-commuting observables. This quantum feature called uncertainty relation 
was first introduced by Heisenberg \cite{HUR}, and then extended by 
Robertson \cite{Robert} for more general observables. The lower bound of the Heisenberg 
uncertainty relation is state dependent.
Later, the uncertainty relation was recast in an entropic form where 
the uncertainty is  measured by the Shannon entropy \cite{SEntropy}. A form 
of the entropic uncertainty  relation (EUR) introduced by Deutsch \cite{EUR1}, 
was subsequently  improved  in the version conjectured in  Ref.\cite{EUR2} and then 
proved in Ref.\cite{EUR3}. (See, Ref.\cite{winweh} for a review of the 
development of EURs).

In the derivation of the above mentioned uncertainty relations, the correlation
of the observed system with another system called quantum memory is 
not  considered. Berta et al., in the Ref.~\cite{QMemory}, discussed the 
possibility of reduction of the lower bound of EUR in the 
scenario when one considers the correlation of the observed system 
with the quantum memory. For example, when the observed system is maximally 
entangled with the quantum memory, the lower bound of EUR becomes 
zero for the measurement of two non-commuting observables.
This phenomena 
has been brought out in two recent experiments using respectively, pure~\cite{Expt_P} 
and mixed states~\cite{Expt_M}. It has been shown in Ref.~\cite{Pramanik}, that 
 the lower bound of EUR in the presence of quantum memory may be  {\it optimized}
in an experimental scenario using the {\it fine-grained} uncertainty relation (FUR)~\cite{FUR}.
Recently, Coles and Piani \cite{CP_2014} have  developed the analysis in order to
make the bound tighter. 

In the Ref.~\cite{QMemory}, the  authors showed that entanglement is the resource 
to reduce the uncertainty. In a subsequent work, Pati et al.~\cite{Pati}
have claimed that quantum discord~\cite{Cinf1, Cinf2} acts as a resource
when the shared state between quantum memory and observed system is chosen 
from a class of states including Werner states and isotropic states.  However, 
the above resources fail to reduce uncertainty optimally~\cite{Pramanik}, in general.
The motivation of the present work is to find out the physical resources responsible 
for the optimal reduction of entropic uncertainty, which is given operationally by 
using the fine-grained uncertainty relation~\cite{FUR}. 
In other words,  we investigate the question as to 
which physical quantity is responsible for reduction of the uncertainty  
optimally in an experimental situation involving the measurement of two 
incompatible observables in the presence of shared states (correlations) between
two parties.

In the present
work we introduce the measure {\it extractable classical information} which
as we show, contributes exactly to reducing the uncertainty by an amount
leading to the optimal lower bound for several examples of entangled, 
separable as well
as classical states. Here, we derive a new uncertainty relation 
in terms of the  extractable classical information.  
We show that the lower bound of the uncertainty relation
derived here is equal to the optimal lower bound  obtained
with the help of the fine-grained uncertainty relation for various
pure and mixed states.
It further follows from our  relation  
that even in the 
absence of quantum correlations between the two parties, the  
uncertainty may be reduced with the help of classical correlations.

\section{Definitions and Mathematical Preliminaries}

The entroipc form of uncertainty relation, when the correlation of the 
observed system with quantum memory is not considered, is given 
by~\cite{EUR2,EUR3}
\begin{eqnarray}
\mathcal{H}(R)+\mathcal{H}(S) \geq\log_2 \frac{1}{c},
\label{EUR1}
\end{eqnarray}
where  $\mathcal{H}(k) =- \sum_i p_i^k \log_2 p_i^k$  is the  Shannon entropy with $p_i^k$
being the probabiality of the $i$-th outcome for the measurement of observable 
$k~\in\{R,S\}$). The 
complementarity of the observables $R$ and $S$ is measured by the quantity $ c $~($= \max_{i,j} |\langle r_i|s_j\rangle|^2$, with $|r_i\rangle$ 
and $|s_j\rangle$ are  the eigenvectors of $R$ and $S$, respectively).

To discuss the EUR in the presence of quantum memory, one may consider the 
following game as discussed in the Ref.~\cite{QMemory}. Bob prepares
two system $A$ and $B$ in a bipartite quantum state $\rho_{AB}$ and
sends the system $A$ to Alice. Now, Alice is going to  measure either 
observable $R$ or $S$ on her system $A$. From the knowledge of the 
system $B$, Bob's task is to miniize his uncertainty about Alice's 
measurement outcome. Bob is able to reduce his uncertainty
about Alice's measurement outcome with the help of communication from Alice
regarding the choice of her
measurement performed, but not its outcome. The modified form of EUR in the presence 
of quantum memory is given by \cite{CP_2014}
\begin{eqnarray}
\mathcal{S}(R_A|B)+\mathcal{S}(S_A|B) \geq c^{\prime}(\rho_{A}) + \mathcal{S}(A|B)
\label{EUR-QM}
\end{eqnarray}
where $\mathcal{S}(A|B)$ ($=\mathcal{S}(\rho_{AB})-\mathcal{S}(\rho_B)$, where $\rho_B=Tr_{A}[\rho_{AB}]$)  is called the 
conditional von-neumann entropy of the state $\rho_{AB}$ and $c^{\prime}(\rho_{A}) = \max\{c^{\prime}(\rho_A,R_A,S_A), c^{\prime}(\rho_A,S_A,R_A) \}$. $c^{\prime}(\rho_A,R_A,S_A)$ and $c^{\prime}(\rho_A,S_A,R_A)$ are defined by
\begin{eqnarray}
c^{\prime}(\rho_A,R_A,S_A) = \displaystyle\sum_i p^r_i \log_2 \frac{1}{\max_j c_{ij}} \nonumber \\
c^{\prime}(\rho_A,S_A,R_A) = \displaystyle\sum_j p^s_j \log_2 \frac{1}{\max_i c_{ij}}, 
\label{Complemetary_RS}
\end{eqnarray}
where $p^r_i = \langle r |\rho_{A}|r \rangle$ with $\sum_i p^r_i = 1$,
$p^s_j =\langle s |\rho_{A}|s \rangle $ with $\sum_j p^s_j=1$ and $c_{i,j} = |\langle r_i|s_j\rangle|^2$, i.e., overlap between eigenvector of the observables $R$ and $S$.
Here, the uncertainty for the measurement of the observable $R_A$ ($S_A$) on 
Alice's system by accessing the information stored in the 
quantum memory with Bob  is measured by $\mathcal{S}(R_A|B)$ ($\mathcal{S}(S_A|B)$) 
which is the conditional von Neumann entropy of the  state given by 
\begin{eqnarray}
\rho_{R_A(S_A)B}&=&\sum_{j} (|\psi_j\rangle_{R_A(S_A)}\langle\psi_j|\otimes I)\rho_{AB}(|\psi_j\rangle_{R_A(S_A)}\langle\psi_j|\otimes I)\nonumber \\
&=&\sum_j p_j^{R_A(S_A)} \Pi_j^{R_A(S_A)}\otimes \rho_{B|j}^{R_A(S_A)},
\label{QState}
\end{eqnarray}
where $\Pi_j^{R_A(S_A)}$'s are the orthogonal projectors on the eigenstate $|\psi_j\rangle_{R_A(S_A)}$ of observable $R_A  (S_A)$, $p_j^{R_A(S_A)}=Tr[(|\psi_j\rangle_{R_A(S_A)}\langle\psi_j|\otimes I)\rho_{AB}(|\psi_j\rangle_{R_A(S_A)}\langle\psi_j|\otimes I)]$, $\rho_{B|j}^{R_A(S_A)}=Tr_A[(|\psi_j\rangle_{R_A(S_A)}\langle\psi_j|\otimes I)\rho_{AB}(|\psi_j\rangle_{R(S)}\langle\psi_j|\otimes I)]/p_j^{R_A(S_A)}$ and $\rho_{AB}$ is the state of joint system `$A$' and `$B$'. 
 EUR in presence of quantum memory is modified by the quantity 
$\mathcal{S}(A|B)$ which measures the amount of one-way distillable 
entanglement \cite{DisEnt}. For shared maximal entanglement (i.e., $S(A|B)=-1$)
 between the system and the memory, there is no uncertainty in the measurement 
of incompatible observables. EUR in the presence of quantum memory has been
brought out in two recent experiments using respectively, pure \cite{Expt_P} 
and mixed states \cite{Expt_M}. For experimental purposes \cite{Expt_M}, one can obtain the 
 uncertanty relation form the inequality (\ref{EUR-QM}) with the help of the
relation~\cite{Expt_M}
\begin{eqnarray}
\mathcal{H}(p_d^R) + \mathcal{H}(p_d^S) \geq \mathcal{S}(R_A|B)+\mathcal{S}(S_A|B),
\label{fano_Rel}
\end{eqnarray}
and it is given by
\begin{eqnarray}
\mathcal{H}(p_d^R) + \mathcal{H}(p_d^S) \geq c^{\prime}(\rho_{A}) + \mathcal{S}(A|B),
\label{EUR-QM_Exp}
\end{eqnarray}
where $p_d^R$ ($p_d^S$) is the probability of getting different outcomes when 
Alice and Bob measure the same observables $R$ ($S$) on their respective 
system. Here the lower bound of the sum of uncertainties for the shared 
state $\rho_{AB}$ is given by 
\begin{eqnarray}
\mathcal{L}_1(\rho_{AB})=c^{\prime}(\rho_{A}) + \mathcal{S}(A|B)
\label{LB_1}
\end{eqnarray}

Later, Pati et al. \cite{Pati} have derived a tighter lower bound
 of the uncertainty relation using the state $\rho_{R_A(S_A)B}$. 
 One can improve the above lower bound using inequality (\ref{EUR-QM})  
and it is given by  Eq.(\ref{QState}) to be 
\begin{eqnarray}
\mathcal{S}(R_A|B)+\mathcal{S}(S_A|B) \geq  c^{\prime}(\rho_{A})  + \mathcal{S}(A|B)  + \max\{0,D_A(\rho_{AB})-C_A^M(\rho_{AB})\}, 
\label{EUR-QM_P}
\end{eqnarray}
where the quantum discord $D_A(\rho_{AB})$ is given by  \cite{Cinf1, Cinf2}
\begin{eqnarray}
D_A(\rho_{AB}) = \mathcal{I}(\rho_{AB})-  C_A^M(\rho_{AB}),
\label{QDis}
\end{eqnarray}
with $\mathcal{I}(\rho_{AB})$ ($=\mathcal{S}(\rho_A)+\mathcal{S}(\rho_B)-\mathcal{S(\rho_{AB})}$) being the mutual information of the state $\rho_{AB}$ which 
contains the total correlation present in the state $\rho_{AB}$ shared between 
the system $A$ and the system $B$, and the 
classical information $C_A^M(\rho_{AB})$ for the shared state $\rho_{AB}$ (when Alice measures on her system) is given by
\begin{eqnarray}
C_A^M(\rho_{AB}) = \max_{\Pi^{R_A}}[\mathcal{S}(\rho_B) - \displaystyle\sum_{j} p_j^{R_A} \mathcal{S}(\rho_{B|j}^{R_A}) ] 
\label{Cinf_M}
\end{eqnarray}
In this case, the lower bound of the sum of Bob's uncertainty about Alice's measurement outcome for the measurement of observable $R$ and $S$ is given by
\begin{eqnarray}
\mathcal{L}_2(\rho_{AB}) = c^{\prime}(\rho_{A}) + \mathcal{S}(A|B) + \max\{0,D_A(\rho_{AB})-C_A^M(\rho_{AB})\},
\label{LB_2}
\end{eqnarray}
which becomes a tighter lower bound compared to $\mathcal{L}_1$ given by 
Eq.(\ref{LB_1}) for those state whose quantum discord is larger than the 
classical information, which is true for example,  for a class 
of states including Werner states and isotropic states.

A new form of the uncertainty relation, {\it viz.},
{\it fine grained uncertainty relation}, was proposed by Oppenheim and Wehner \cite{FUR},  motivated by the realization that entropic
functions provide a rather coarse way of
measuring the uncertainty of a set of measurements,
as they do not distinguish the uncertainty
inherent in obtaining any combination of outcomes
 for different measurements. 
 In particular, they considered
a game according to which Alice and Bob both receive binary questions,
i.e., projective spin measurements along two different directions at each side.
The winning probability is given by the
relation \cite{FUR}
\begin{eqnarray}
P^{game}(\mathcal{T}_A,\mathcal{T}_B,\rho_{AB})= \displaystyle\sum_{t_A,t_B} p(t_A,t_B) \displaystyle\sum_{a,b} V(a,b|t_A,t_B) \langle (A_{t_A}^a\otimes B_{t_B}^b) \rangle_{\rho_{AB}}
&\leq & P^{game}_{max}  
\label{FUR1}
\end{eqnarray}
where $\rho_{AB}$ is a  bipartite state shared by  Alice and Bob, and $\mathcal{T}_A$ and $\mathcal{T}_B$ represent the set of measurement settings $\{t_A\}$ and $\{t_B\}$ chosen by Alice and Bob, respectively, with probability $p(t_A,t_B)$.
Alice's (Bob's) question and answer are $t_A (t_B)$ and $a (b)$,
respectively,  with
$A_{t_A}^a=\frac{[I+(-1)^a A_{t_A}]}{2}$ ($B_{t_B}^b=\frac{[I+(-1)^b B_{t_B}]}{2}$)
being a measurement of the observable $A_{t_A}$ ($B_{t_B}$). Here
$V(a,b|t_A,t_B)$ is some function determining the winning condition of the
game, which corresponding to  a special class of nonlocal retrieval games
(CHSH game \cite{FUR}) for which there exist only
one winning answer for one of the two parties, is given by $V(a,b|t_A,t_B)=1$, iff $a\oplus b=t_A.t_B$, and
 $0$ otherwise. $P^{game}_{max}$ is the maximum winning probability of the game, 
maximized over the set of projective spin measurement settings $\{t_A\}$ ($\in$ $\mathcal{T}_A$) by Alice, the set of projective spin measurement settings $\{t_B\}$ ($\in$ $\mathcal{T}_B$) by Bob,
i.e.,
$P^{game}_{max}=\max_{\mathcal{T}_A,\mathcal{T}_B,\rho_{AB}} P^{game}(\mathcal{T}_A,\mathcal{T}_B,\rho_{AB})$.
Using the maximum winning probability it is possible to
discriminate between classical theory, quantum theory and no-signaling
theory with the
help of the degree of nonlocality \cite{FUR}.
A generalization
to the case of tripartite systems has also been proposed \cite{FUR2}.

In a recent work \cite{Pramanik}, we have shown that the lower bound of 
the uncertainty relation given by Eqs. (\ref{EUR-QM}) and (\ref{EUR-QM_P}) are 
 not {\it optimal} (for the choices of their observables that maximally reduce Bob's uncertainty about Alice's measurement outcome), as illustrated by the analysis of
an experiment using mixed states \cite{Expt_M}. We have obtained the {\it optimal} lower bound of entropic uncertainty using fine-grained uncertainty 
relation \cite{FUR}.
Considering a situation \cite{Expt_M} where Alice and Bob both measure 
the same observable on their system, we have derived a new uncertainty relation
 that captures the optimal lower bound for Bob's uncertainty about Alice's 
measurement outcomes. Our uncertainty relation is given by \cite{Pramanik}
\begin{eqnarray}
\mathcal{H}(p_d^R) + \mathcal{H}(p_d^S) \geq \mathcal{H}(p_d^{\sigma_z}) + \mathcal{H}(p_{\inf}^S),
\label{FUR_Opti}
\end{eqnarray}
where the lower bound given by
\begin{eqnarray}
\mathcal{L}_3(\rho_{AB}) = \mathcal{H}(p_d^{\sigma_z}) + \mathcal{H}(p_{\inf}^S)
\label{LB_3}
\end{eqnarray}
is optimal if it is tight since for each pair of observables $R, S$ there is a 
state for which we get the equality. $\mathcal{L}_3(\rho_{AB})$ is obtained with the help 
of the fine-grained uncertainty relation \cite{FUR} which gives the infimum 
winning probability $p_d^{\sigma_z}$ ($p_{\inf}^S$) (corresponding to the minimum
uncertainty) for the measurement of 
observable $\sigma_z$ ($\sigma_S=\vec{n}_S.\vec{\sigma}\neq \sigma_z$  with $\vec{n}_S\equiv\{\sin(\theta_S)\cos(\phi_S),\sin(\theta_S)\sin(\phi_S),\cos(\theta_S)\}$ being a unit vector and $\vec{\sigma}=\{\sigma_x,\sigma_y,\sigma_z\}$ are the Pauli matrices) corresponding to the game ruled by the  winning condition 
given by \cite{Pramanik}
\begin{eqnarray}
V(a,b) 	&=& 1 \hspace{1cm} \text{iff}~a\oplus b=1, \nonumber \\
		&=& 0 \hspace{1cm} \text{otherwise},
\end{eqnarray}
where `$a$' and `$b$' are the binary outcomes (i.e., $\{a,~b\}\in \{0,1\}$) for Alice and Bob, respectively. $V(a,b)$ describes the experimental situation used 
in the Ref.\cite{Expt_M}, i.e., Alice and Bob measure the same observables and 
calculate the probability of getting different outcomes. In measurements and 
communication involving two parties, the lower bound of entropic
uncertainty cannot fall below the bound (\ref{LB_3}).
Using the same approach of Ref.~\cite{Pramanik}, a fine-grained steering inequality has been derived~\cite{Fur_Steer} which provides the most optimal steering condition for two qubit systems.

\section{Uncertainty relation using extractable classical information}

To derive the sum of uncertainties for the measurement of two incompatible 
observables $R$ and $S$, we consider the following memory game \cite{QMemory}.
 In this game Bob prepares a particle (labeled by `$A$') in a particular 
state, say, $\rho_A$ and sends it to Alice who measures an observable chosen 
from the non-commuting set $\{R_A,S_A\}$ and communicates only the choice of 
the observable to Bob. Bob's task is to reduce his uncertainty about the 
Alice's measurement outcome. To win the game, Bob chooses one of the following 
two strategies -- (i) {\it classical strategy}; (ii) {\it quantum strategy}.

{\it Classical strategy : } Here, Bob prepares two particles (say, 1st particle labeled by $A$ and 2nd particles labeled by $B$) in the identical 
state, $\rho=\rho_A=\rho_B$. The combined state of two particles is given by
\begin{eqnarray}
\rho_{AB}=\rho_A \otimes \rho_B.
\label{PState_QM}
\end{eqnarray}
After preparation, Bob sends the 1st particle to Alice. When Alice communicates
 about her choice of measurement from the set of observables $\{R_A,S_A\}$, 
Bob measures the same observable on the $2$nd particle possessed by himself. 
He infers about the Alice's measurement outcome from his own measurement 
outcome. Note here that Bob keeps full information of the state of Alice
since he himself has prepared it.
Here, the uncertainty relation prevents Bob to know 
with arbitrary precision the measurement outcomes of two non-commuting 
observables.
 The EUR which gives the lower bound for the measurement of the above two 
non-commuting observables follows from Eq.(\ref{EUR-QM}) for product states
and is given by \cite{CP_2014, QMemory2}
\begin{eqnarray}
\mathcal{H}(R_B)+\mathcal{H}(S_B) \geq c^\prime (\rho_{B}) + \mathcal{S}(\rho_B),
\label{C_Strategy_QM}
\end{eqnarray}
where the subscript $B$ labels Bob's measurement. The inequality (\ref{C_Strategy_QM}) is tighter  than the entropic
uncertainty relation given by inequality (2), and hence, Bob can not reduce his
 uncertainty about Alice's measurement outcome below the lower bound $\mathcal{L}_0(\rho_{AB})$ given by
\begin{eqnarray}
\mathcal{L}_0(\rho_{AB})=c^\prime (\rho_{B}) + \mathcal{S}(\rho_B).
\label{LB_0}
\end{eqnarray} 
Note that, the state given by Eq.(\ref{PState_QM}) has zero classical correlation (i.e., $C_A^M=0$) and zero quantum correlation (i.e., $D_A=0$) \cite{ZDiscord}. The inequality (\ref{C_Strategy_QM}) represents the entropic
uncertainty relation for Bob's measurements of two non-commuting
observables $R_B$ and $S_B$  on his system,  and pertains to the situation
when there is either no correlation with the other
system called quantum memory, or the correlation with the quantum memory is not
considered.

{\it Quantum strategy : } In this strategy, Bob prepares two particles in a
correlated state, $\rho_{AB}$, and sends the $1$st particle to Alice and keeps 
the $2$nd particle. To reduce his uncertainty further from the bound $c^\prime (\rho_{B}) + \mathcal{S}(\rho_B)$ (which is the lower bound of uncertainty  corresponding to 
the {\it classical strategy}), Bob uses the correlations 
(quantum and/or classical) present in the state $\rho_{AB}$. After getting 
information about the choice of measurement, Bob measures the same observable 
as Alice's choice. Since, Alice and Bob measure independently on their 
respective systems,  the order of measurement, i.e., who measures first, does 
not affect in the considered game. Here consider Alice communicates about her choice of observable from the set $\{R,S\}$ to Bob. First Bob measures the observable and then Alice measures. After the measurement performed by Bob with the observable that communicated by Alice, the combined state $\rho_{AR_B(S_B)}$ is given by 
\begin{eqnarray}
\rho_{AR_B(S_B)}= \sum_j p_j^{R_B(S_B)} \rho_{A|j}^{R_B(S_B)} \otimes  \Pi_j^{R_B(S_B)},
\label{QState_2}
\end{eqnarray}
where $\Pi_j^{R_B(S_B)}=|\psi\rangle_{R_B(S_B)}\langle\psi |$, $\rho_{A|j}^{R_B(S_B)}$ ($=Tr_B[(I \otimes \Pi_j^{R_B(S_B)})\rho_{AB}]$) is the Alice's conditional state when Bob gets $j$-th outcome and $p_j^{R_B(S_B)}$ $(=Tr[(I \otimes |\psi\rangle_{R_B(S_B)}\langle\psi |)\rho_{AB}])$ is the probability of getting $j$-th outcome by Bob. 

The classical information $C_B^M(\rho_{AB})$, given by
\begin{eqnarray}
C_B^M (\rho_{AB}) = \max_{\Pi^{R_B}}[\mathcal{S}(\rho_A) - \displaystyle\sum_{j} p_j^{R_B} \mathcal{S}(\rho_{A|j}^{R_B}) ],
\label{Cinf_M_B}
\end{eqnarray} 
where $\rho_A=Tr_{B}[\rho_{AB}]$, gives the {\it maximum information} that Bob can extract on average about the Alice's system by measuring on his system when they share the state $\rho_{AB}$. Now, one may ask the following questions -- what information can Bob extract about Alice's measurement outcomes? 
$C_B^M(\rho_{AB})$ contains the information about the Alice's measurement 
outcomes when she measures along a particular direction which maximizes the 
quantity $C_B(\rho_{AB})$ (where $C_B(\rho_{AB})$ is taken with out maximization 
in Eq.(\ref{Cinf_M_B})).  When Bob gets the $j$-th outcome for the measurement 
of the observable $R_B$ on his system, his knowledge about Alice's measurement 
outcomes for the measurement in the eigenbasis of $\rho_{A|j}^{R_B}$ is given by 
the quantity $\mathcal{S}(\rho_{A|j}^{R_B})$. Since $\mathcal{S}(\rho_A)$
is Bob's uncertainty about Alice's outcome in the absence of correlations,
from the Eq.(\ref{Cinf_M_B}), it can be easily seen that $C_B^M(\rho_{AB})$  
measures the amount of  Bob's uncertainty about Alice's measurement outcome 
reduced due to Bob's measurement. 
For example, for the shared Werner state $\rho_{AB}^W$ \cite{WState} between Alice and Bob given by
\begin{eqnarray}
\rho_{AB}^W=\frac{1-p}{4} I\otimes I + p |\psi^-\rangle\langle\psi^-|,
\label{rhoW}
\end{eqnarray}
where $I$ is the ($2\otimes 2$) unitary matrix, $|\psi^-\rangle$ is the singlet state $(|01\rangle_{AB}-|10\rangle_{AB})/\sqrt{2}$, and $p$, the mixedness parameter (lying between $0$ and $1$),  Bob gets the maximum information 
about Alice's 
measurement outcomes given by $\mathcal{S}(\rho_{A|j}^{R_B}) $ when they measure 
the same observables on their respective system. Hence, classical information 
quantifies Bob's maximum knowledge about Alice's measurement outcome in a 
specific direction, say in the eigen basis of $\rho_{A|j}^{R_B}$.

According to our considered game, when Alice communicates her choice, say, 
$R_A$ (where `A' labels Alice's choice),  Bob measures same the observable 
$R_B=R_A$ on his particle (labeled by `$B$'). Due to Bob's measurement, the
reduced uncertainty measured by the conditional von-Neumann entropy of the 
state, $\rho_{AR_B}$ given by Eq.(\ref{QState_2}) now becomes
\begin{eqnarray}
\mathcal{S}(A|R_B) =
 \mathcal{S}(\rho_A) - C_B^R (\rho_{AB}),
\label{LHS1}
\end{eqnarray}
where $C_B^R (\rho_{AB}) =\mathcal{S}(\rho_A) - \sum_i p_{i}^{R_B} \mathcal{S}(\rho_{A|i}^{R_B})$ as obtained from the Eq.(\ref{Cinf_M_B}) without taking the
maximization. This is the information obtained by Bob 
when he 
makes a measurement of the observable $R_B$ on his system. 
$C_B^R (\rho_{AB})$ gives the information about Alice's measurement outcomes 
when she measures in the eigenbasis of $\rho_{A|i}^{R_B}$ on her particle. Bob's 
maximum information about Alice's measurement outcome in the eigenbasis 
$\rho_{A|i}^{R_B}$ is given by 
$C_B^M(\rho_{AB})=\max_{R_B}C_B^R (\rho_{AB})$
which is known as the classical information where the maximization is taken 
over all possible observables $R_B$. After Bob's measurement, Alice measures 
the observable $R_A$ on her particle and the combined state~(\ref{QState_2})
becomes
\begin{eqnarray}
\rho_{R_A,R_B\,(S_A,S_B)} = \sum_i p_i^{R_B(S_B)} \left( \sum_k q_{k|i}^{R_A(S_A)}  \Pi_{k}^{R_A(S_A)} \right) \otimes  \Pi_i^{R_B(S_B)},
\label{QState_222}
\end{eqnarray}
where $\Pi_{k}^{R_A\,(S_A)}$ is projector corresponding to the eigenstate of 
observable $R_A$ ($S_A$) and $q_{k|i}^{R_A\,(S_A)}=Tr[\Pi_{k}^{R_A\,(S_A)}\; \rho_{A|i}^{R_B\,(S_B)}]$ is the conditional probability distribution for the measurement of observable $R_A$ ($S_A$) on Alice's particle, given that Bob gets $i$th outcome for the measurement of the same observable $R_B$ ($S_B$) on his particle. Now, Alice's reduced uncertainty for the 
measurement of observable $R_A$, i.e., more specificaly, conditional entropy of the
state $\rho_{R_A,R_B}$ is given by 
\begin{eqnarray}
\mathcal{H}(R_A|R_B) = \mathcal{H}(R_A) - C_{A,B}^{R,R}(\rho_{AB}),
\label{LHS11}
\end{eqnarray}
with
\begin{eqnarray}
C_{A,B}^{R,R} (\rho_{AB})=\mathcal{H}(R_A) - \sum_i p_{i}^{R_B} \mathcal{H}(q^{R_A}_{i}),
\label{CinfRR}
\end{eqnarray}
where $\mathcal{H}(R_A)$ is the Shannon entropy of the probability distribution $\{q_{k}^{R_B}\}$ corresponding to different measurement outcomes $\{k\}$ for the measurement of observable $R_A$ on Alice's particle and $\mathcal{H}(q^{R_A}_{i})$ is the Shannon entropy of the conditional probability distribution $\{q_{k|i}^{R_A}\}$.  
We define the quantity  $C_{A,B}^{R,R} (\rho_{AB})$  as the 
 ``extractable classical information''.

Similarly, when both Alice and Bob measures the observable $S$, the conditional entropy of the state $\rho_{S_A,S_B}$ (given by Eq.~(\ref{QState_222})) becomes
\begin{eqnarray}
\mathcal{H}(S_A|S_B) = \mathcal{H}(S_A) - C_{A,B}^{S,S} (\rho_{AB}),
\label{LHS22}
\end{eqnarray}
where $C_{A,B}^{S,S} (\rho_{AB})$ is the extractable classical information for the measurement of the observable $S$ on the both particles. Now, combining
  Eqs. (\ref{LHS11}) and (\ref{LHS22}), one gets
\begin{eqnarray}
\mathcal{H}(R_S|R_B) + \mathcal{H}(S_A|S_B) = \mathcal{H}(R_A) + \mathcal{H}(S_A) - C_{A,B}^{R,R} (\rho_{AB}) - C_{A,B}^{S,S} (\rho_{AB})
\label{intermed}
\end{eqnarray}
The sum of the first two terms on the r.h.s. of the above equation 
(\ref{intermed}) represents the entropy of a single system (system $A$)
when there is either no correlation with the other
system called quantum memory, or the correlation with the quantum memory is not
considered. Hence, the sum $\mathcal{H}(R_A) + \mathcal{H}(S_A)$ can be
contrained through the
inequality (\ref{C_Strategy_QM}), using which we
obtain 
\begin{eqnarray}
\mathcal{H}(R_A|R_B)+\mathcal{H}(S_A|S_B) \geq  c^\prime (\rho_{A})+ \mathcal{S}(\rho_A) - C_{A,B}^{R,R}(\rho_{AB})- C_{A,B}^{S,S} (\rho_{AB}),
\label{EUR-QM41}
\end{eqnarray}
where $\rho_A$ is the density state of Alice's particle. Now, using the 
inequality~(\ref{fano_Rel}), Eq.({\ref{EUR-QM41}) becomes
\begin{eqnarray}
\mathcal{H}(p^R_d) + \mathcal{H}(p^S_d) \geq  c^\prime (\rho_{A})+ \mathcal{S}(\rho_A) - C_{A,B}^{R,R}(\rho_{AB}) - C_{A,B}^{S,S} (\rho_{AB}),
\label{EUR-QM4}
\end{eqnarray}
where $\mathcal{H}(p^\alpha_d)$ is the Shannon entropy of the probability distribution $\{p^\alpha_d\}$ when Alice and Bob measure same observable $\alpha\in\{R,S\}$ and get different outcomes. Eq.(\ref{EUR-QM4}) represents our new 
uncertainty relation when both Alice and Bob measure two incompatible
observables $R$ and $S$. Hence, the lower bound of Bob's uncertainty about 
Alice's measurement outcomes is given by
\begin{eqnarray}
\mathcal{L}_4(\rho_{AB})=  c^\prime (\rho_{A})+ \mathcal{S}(\rho_A) - C_{A,B}^{R,R}(\rho_{AB})  - C_{A,B}^{S,S} (\rho_{AB}).
\label{LB_4}
\end{eqnarray}

\section{Examples}

In the following analysis we compare the bound $\mathcal{L}_4(\rho_{AB})$
 with the lower bounds through the quantum strategy obtained earlier in the 
literature, {\it viz.}, 
 $\mathcal{L}_1(\rho_{AB})$ (given by Eq.(\ref{LB_1})) \cite{QMemory, Expt_M},
the bound $\mathcal{L}_2(\rho_{AB})$ (given by Eq.(\ref{LB_2}) \cite{Pati},
 the bound $\mathcal{L}_3(\rho_{AB})$ (given by Eq.(\ref{LB_3})) \cite{Pramanik},
 as well as
the bound $\mathcal{L}_0(\rho_{AB})$ (given by Eq.(\ref{LB_0}) with 
$\mathcal{S}(\rho_B) = \mathcal{S}(Tr_A\rho_{AB})$) obtained with 
the help of the classical strategy 
 for various classes of pure and mixed entangled and separable
states.  We show that the lower bound given by Eq.(\ref{LB_4})  is 
optimal as obtained through fine-graining \cite{Pramanik} for all the cases 
considered here.

{\bf Pure entangled state :} Here we consider a pure entangled state $\rho_{AB}^{PE}$, given by
\begin{eqnarray}
\rho_{AB}^{PE}=\sqrt{\alpha}|01\rangle_{AB} - \sqrt{1-\alpha} |10\rangle_{AB},
\end{eqnarray}
where $\alpha$ lies between $0$ and $1$, and the state $\rho_{AB}^{PE}$ is 
maximally entangled for $\alpha=\frac{1}{2}$. The  classical information 
(when Alice measures her particle) is given by
\begin{eqnarray}
C_B^M(\rho_{AB}^{PE})=\mathcal{H}(\alpha),
\end{eqnarray}
where $\mathcal{H}(\alpha)= \alpha \log_2\alpha - (1-\alpha)\log_2(1-\alpha)$. $C_B^M(\rho_{AB}^{PE})$ gives the information about Alice's measurement outcome 
in the direction $\{\mu \cos[\phi_S] \sin[\theta_S], \mu \sin[\phi_S] \sin[\theta_S], \frac{1-2\alpha+\cos[\theta_S]}{1+\cos[\theta_S]-2\alpha \cos[\theta_S]}    \}$ (where $\mu =\frac{2\sqrt{\alpha(1-\alpha)}}{1+\cos[\theta_S]-2\alpha \cos[\theta_S]} $) to Bob  when he measures along $\{\sin(\theta_S)\cos(\phi_S),\sin(\theta_S)\sin(\phi_S),\cos(\theta_S)\}$.   Let us consider that before playing 
the game Alice and 
Bob discuss about their strategy, such as, choices of the state and measurement
 settings. Alice chooses those settings for which Bob's uncertainty about her 
measurement outcome will be minimum as well as maximize the lower bound of 
uncertainty in the classical strategy (given by Eq.(\ref{LB_0})), i.e., $c^\prime (\rho_{B})=1$ where $\max_i c_{ij}=\max_j c_{ij} =1/2$. With the help of the fine-grained uncertainty relation 
\cite{Pramanik,FUR}, one can obtain the winning probability (corresponding to
minimum uncertainty) when Alice and Bob 
both measure the same observable  and get different outcomes, i.e., 
$a\oplus b=1$ \cite{Pramanik}. The winning probability is given by
\begin{eqnarray}
P^{\text{game}}(\rho_{AB}^{PE}) = \frac{1}{4} (3+2\sqrt{\alpha (1-\alpha)} + (1-2\sqrt{\alpha(1-\alpha)})\cos[2\theta_S] )
\label{FUR_game_PE}
\end{eqnarray}
Bob's uncertainty about Alice's outcome would be minimum for the choice of observables given by
\begin{eqnarray}
R&=&\sigma_z \hspace{1cm} \text{(i.e.,}~~\theta_S=0), \nonumber \\
S&=&\sigma_x \hspace{1cm} \text{(i.e.,}~~\theta_S=\frac{\pi}{2}),
\label{Sett.}
\end{eqnarray}
which leads to $p_d^{R}=1$, and the infimum probability $p_{\inf}^{S} = 1/2 + \sqrt{\alpha(1-\alpha)}$. Hence, the optimal lower bound obtained from the Eq.(\ref{LB_3}) is given by \cite{Pramanik}
\begin{eqnarray}
\mathcal{L}_3(\rho_{AB}^{PE})=\mathcal{H}(\frac{1}{2}-\sqrt{\alpha (1-\alpha)}).
\label{LB_3_PE}
\end{eqnarray}

When Bob chooses the {\it classical strategy}, he first prepares two copies of the state $Tr_B(\rho_{AB}^{PE})$ and send one to Alice. For the above choice of 
observables, Bob's uncertainty about Alice's measurement outcome is 
maximally reduced ($c^\prime (\rho_B)=1$ by choosing the above measurement settings), and is given by the inequality (\ref{C_Strategy_QM}). The lower bound
 (\ref{LB_0}) is given by 
\begin{eqnarray}
\mathcal{L}_0(\rho_{AB}^{PE})=1+\mathcal{H}(\alpha).
\label{LB_0_PE}
\end{eqnarray}

In the {\it quantum strategy} using the uncertainty relations proposed
in Refs. \cite{QMemory} and \cite{Pati}, Bob's uncertainty is lower bounded  
by  ((\ref{LB_1}) and (\ref{LB_2})), respectively, which turn out to be
equal, given by
\begin{eqnarray}
\mathcal{L}_1(\rho_{AB}^{PE})=\mathcal{L}_2(\rho_{AB}^{PE})=1-\mathcal{H}(\alpha).
\label{LB_12_PE}
\end{eqnarray}
However, in practice Bob is unable to reduce his uncertainty upto the above
level, since  
$\mathcal{L}_1(\rho_{AB}^{PE})$ is not the optimal lower bound. The main reason 
is that Bob only extracts the information $C_{A,B}^{\sigma_z,\sigma_z}(\rho_{AB}^{PE})$ ($C_{A,B}^{\sigma_x,\sigma_x}(\rho_{AB}^{PE})$) given by $\mathcal{H}(\alpha)$ $(1-\mathcal{H}(\frac{1}{2}-\sqrt{\alpha (1-\alpha)}))$ when both of them measure
the same spin observables $R=\sigma_z$ ($S=\sigma_x$) on their respective 
particle. Hence, the lower bound (given by Eq.(\ref{LB_4})) of Bob's 
uncertainty is given by
\begin{eqnarray}
\mathcal{L}_4(\rho_{AB}^{PE})=\mathcal{H}(\frac{1}{2}-\sqrt{\alpha (1-\alpha)}).
\label{LB_4_PE}
\end{eqnarray}
From the Eqs.(\ref{LB_3_PE}) and (\ref{LB_4_PE}), it is clear that the 
quantities $C_{A,B}^{\sigma_z,\sigma_z}(\rho_{AB}^{PE})$ and $C_{A,B}^{\sigma_x,\sigma_x}(\rho_{AB}^{PE})$ are responsible for reducing Bob's uncertainty about Alice's 
measurement outcome optimally.  This explains in terms of physical resources
why the lower bound $\mathcal{L}_1(\rho_{AB}^{PE})$ ($\leq \mathcal{L}_3(\rho_{AB}^{PE})$) is not experimentally reachable, whereas the lower bound $\mathcal{L}_3(\rho_{AB}^{PE})$ given 
by fine-graining is indeed attainable.

\begin{figure}[!ht]
\resizebox{13cm}{8cm}
{\includegraphics{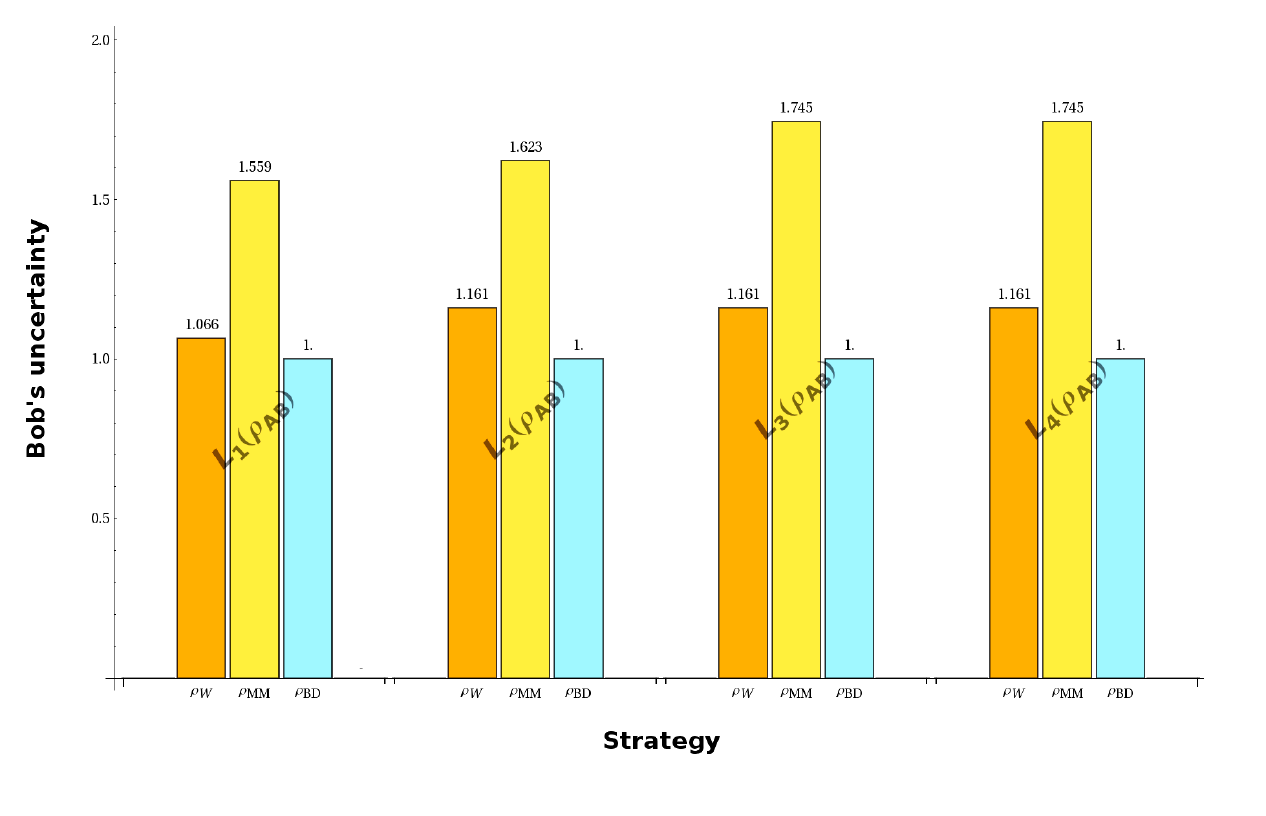}}
\caption{\footnotesize
A comparison of the different lower bounds for the (i) Werner state with
$p=0.723$, (ii) the state with maximally mixed marginals with the
ci's given by $c_x=0.5,~c_y=-0.2,$ and $c_z=-0.3$, and (iii) the Bell diagonal state with 
$p=0.5$.
}
\label{Fig.1}
\end{figure}

{\bf Werner State :} For the class of Werner State $\rho_{AB}^W$, given by Eq.(\ref{rhoW}), the classical information is given by
\begin{eqnarray}
C_B^M(\rho_{AB}^W)=1-\mathcal{H}(\frac{1-p}{2}).
\end{eqnarray}
$C_B^M(\rho_{AB}^W)$ gives Bob the information about the measurement outcome of 
Alice when they measure same observables. The quantum discord  of the state $\rho_{AB}^W$ is given by
\begin{eqnarray}
D_B(\rho_{AB}^W)=\mathcal{I}(\rho_W)-C_B^M,
\end{eqnarray}
where $\mathcal{I}(\rho_{AB}^W)=2+3\frac{1-p}{4}\log_2\frac{1-p}{4}+\frac{1+3 p}{4}\log_2\frac{1+3p}{4}$ is the mutual information of $\rho_{AB}^W$. 

In the {\it Classical strategy}, for the choice observables given by Eq.(\ref{Sett.}) (which minimize Bob's uncertainty optimally \cite{Pramanik}), Bob's uncertainty is lower bounded by (\ref{LB_0})
\begin{eqnarray}
\mathcal{L}_0(\rho_{AB}^{W})=2,
\label{LB_0_W}
\end{eqnarray}
where $\rho_A^{W}=Tr_B[\rho_{AB}^W]=\frac{I}{2}$. When Bob uses the {\it quantum strategy} \cite{QMemory, Expt_M}, his uncertainty (given by Eq.(\ref{EUR-QM_Exp})) is bounded by 
\begin{eqnarray}
\mathcal{L}_1(\rho_{AB}^{W})=2-\mathcal{I}(\rho_{AB}^W),
\label{LB_1_W}
\end{eqnarray}
where for the state $\rho_{AB}^W$, $\mathcal{S}(A|B)=1- \mathcal{I}(\rho_{AB}^W)$.The  improved lower bound (\ref{LB_2}) given by Pati et al.\cite{Pati} for the Werner calss of states  turns out to be 
\begin{eqnarray}
\mathcal{L}_2(\rho_{AB}^W)= 2-2 C_B^M(\rho_{AB}^W)=2 \mathcal{H}(\frac{1-p}{2}).
\label{LB_2_W}
\end{eqnarray}
Note that Bob is able to gain his knowledge about Alice's measurement outcomes by an amount $C_B^M(\rho_{AB}^W)$ when both Alice and Bob measure the same 
observables $R=\sigma_z$ ($S=\sigma_x$) on their respective particles. Hence, 
Bob's uncertainty (given by Eq.(\ref{EUR-QM4})) is lower bounded by   
\begin{eqnarray}
\mathcal{L}_3(\rho_{AB}^{W})=2-2 C_B^M(\rho_{AB}^W)=2 \mathcal{H}(\frac{1-p}{2}).
\label{LB_3_W}
\end{eqnarray} 
Now, using fine-graining the optimal lower bound for Bob's uncertainty is 
given by \cite{Pramanik}
\begin{eqnarray}
\mathcal{L}_4(\rho_{AB}^W)=2 \mathcal{H}(\frac{1-p}{2}),
\label{LB_4_W}
\end{eqnarray}
Thus, for the Werner class of states,  Bob can actually minimize his 
uncertainty about Alice's measurement outcome upto $2 \mathcal{H}(\frac{1-p}{2})$ (given by Eqs.(\ref{LB_2_W}), (\ref{LB_3_W}) and (\ref{LB_4_W})). The lower bound $\mathcal{L}_1(\rho_{AB}^{W})$ ($\leq \mathcal{L}_3(\rho_{AB}^{W})$) is not 
experimentally reachable.

{\bf Bell diagonal state :} The Bell diagonal state, used in  Ref.\cite{Expt_M} is given by
\begin{eqnarray}
\rho_{AB}^{BD}= p \rho_2+(1-p)\rho_S,
\label{QState_BD}
\end{eqnarray}
where $\rho_2$ is the density matrix of the state $\frac{|00\rangle +|11\rangle}{\sqrt{2}}$. The classical information of the state $\rho_{AB}^{BD}$ is given by
\begin{eqnarray}
C_B^M(\rho_{AB}^{BD})=1.
\label{Cinf_M_BD}
\end{eqnarray}
Here $C_B^M(\rho_{AB}^{BD})$ gives Bob the information about Alice's measurement outcome for the measurement along  $\{\sin(\theta_S)\cos(\phi_S),- \sin(\theta_S)\sin(\phi_S),\cos(\theta_S)\}$    from his measurement outcome along the 
direction $\{\sin(\theta_S)\cos(\phi_S),\sin(\theta_S)\sin(\phi_S),\cos(\theta_S)\}$. The quantum discord of $\rho_{AB}^{BD}$ is given by
\begin{eqnarray}
D_B(\rho_{AB}^{BD})=1-\mathcal{H}(p),
\label{QD_BD}
\end{eqnarray}
where $\mathcal{I}(\rho_{AB}^{BD})$ $(=2-\mathcal{H}(p))$ is the mutual 
information of the state $\rho_{AB}^{BD}$. From the Eqs.(\ref{Cinf_M_BD}) and 
(\ref{QD_BD}), it is clear that for the state $\rho_{AB}^{BD}$, $C_B^M(\rho_{AB}^{BD})\geq D_B(\rho_{AB}^{BD})$. 

In the {\it classical strategy} for the choice of above observables, Bob's uncertainty is bounded by 
\begin{eqnarray}
\mathcal{L}_0(\rho_{AB}^{BD})=2.
\label{LB_0_BD}
\end{eqnarray}
In the {\it quantum strategy}, theoretically Bob's uncertainty (obtained using Eq.({\ref{EUR-QM}) and (\ref{EUR-QM_P})) is lower bounded by 
\begin{eqnarray}
\mathcal{L}_1(\rho_{AB}^{BD})=\mathcal{L}_2(\rho_{AB}^{BD})=\mathcal{H}(p),
\label{LB_1_BD}
\end{eqnarray}
where $S(A|B) = \mathcal{H}(p)-1$. With the measurement on his system of an 
observable communicated by Alice, Bob extracts  the classical information by 
an amount $C_{A,B}^{\sigma_z,\sigma_z}(\rho_{AB}^{BD})=1-\mathcal{H}(p)$ ($C_{A,B}^{\sigma_y,\sigma_y}(\rho_{AB}^{BD})=1$) for the spin measurement along $z-$ direction ($y-$ direction). Hence, Bob's uncertainty is lower bounded by
\begin{eqnarray}
\mathcal{L}_4(\rho_{AB}^{BD})= \mathcal{H}(p).
\label{LB_4_BD}
\end{eqnarray}
In this case the optimal lower bound for Bob's uncertainty about Alice's measurement outcome given by fine-graining \cite{Pramanik} also turns out to be
\begin{eqnarray}
\mathcal{L}_3(\rho_{AB}^{BD})=\mathcal{H}(p).
\label{LB_3_BD}
\end{eqnarray}
Here the lower bound predicted by \cite{QMemory,Pati} is optimal. Eqs.(\ref{LB_4_BD}) and (\ref{LB_3_BD}) show that the extactable classical information $C_{A,B}^{\sigma_z,\sigma_z}(\rho_{AB}^{BD})=1-\mathcal{H}(p)$, $C_{A,B}^{\sigma_y,\sigma_y}(\rho_{AB}^{BD})=1$ is responsible for reducing Bob's uncertainty optimally.

{\bf Maximally mixed marginal state :} The maximally mixed marginal state $\rho_{AB}^{MM}$ is given by
\begin{eqnarray}
\rho_{AB}^{MM}= \frac{1}{4}(\mathcal{I}+\sum_{i=x,y,z} c_i\sigma_i\otimes\sigma_i).
\label{QState_MM}
\end{eqnarray}
where the coefficients $c_i$'s ($i\in\{x,y,z\}$) are constrained by the eigenvalues $\lambda_i\in[0,1]$ of $\rho_{AB}^{MM}$ given by
\begin{eqnarray}
\lambda_0=\frac{1-c_x-c_y-c_z}{4}, ~
\lambda_1=\frac{1-c_x+c_y+c_z}{4}, ~
\lambda_2=\frac{1+c_x-c_y+c_z}{4}, ~
\lambda_3=\frac{1+c_x+c_y-c_z}{4}. 
\label{EValue_MM}
\end{eqnarray}
The mutual information of the state $\rho_{AB}^{MM}$ is given by
\begin{eqnarray}
\mathcal{I}(\rho_{AB}^{MM}) = 2+\displaystyle\sum_{j=0}^{3} \lambda_j \log2[\lambda_j].
\label{MI_MM}
\end{eqnarray}
The  classical information of the state is given by \cite{LUO}
\begin{eqnarray}
C_B^M(\rho_{AB}^{MM}) =  \frac{1-c_M}{2} \log2[1-c_M] + \frac{1+c_M}{2} \log2[1+c_M],
\label{Cinf_M_MM}
\end{eqnarray}
where $c_M=\max[|c_x|,|c_y|,|c_z|]$, and the quantum discord of the state $\rho_{AB}^{MM}$ is given by
\begin{eqnarray}
D_B(\rho_{AB}^{MM})=\mathcal{I}(\rho_{AB}^{MM})-C_B^M(\rho_{AB}^{MM}).
\label{D_MM}
\end{eqnarray}
As usual,
before playing the game, Alice and Bob discuss the measurement settings (i.e., 
strategy for the game) for the shared state $\rho_{AB}^{MM}$. To optimize the 
uncertainty, Bob takes the help of the fine-grained uncertainty relation (FUR) 
\cite{Pramanik}. Here, the winning probability when Alice and Bob both
 measure 
the observable $S$ is given by
\begin{eqnarray}
P^{\text{game}}_S=\frac{1}{2} (1 - c_x \sin^2[\theta_S] \cos^2[\phi_S]  - c_y \sin^2[\theta_S] \sin^2[\phi_S] -c_z\cos^2[\theta_S] ).
\end{eqnarray}
The measurement settings may be chosen such that the quantity $P^{\text{game}}_S$ is maximized. For the measurement setting $\sigma_z$ (i.e., $\theta_S=0$), $P^{\text{game}}_{\sigma_z}$ will be maximum when $c_z-c_x<0$.
Here we consider $c_x=0.5,~c_y=-0.2,$ and $c_z=-0.3$, and for this choices, 
the observable $R=\sigma_z$ and $S=\sigma_x$ minimizes Bob's uncertainty
\cite{Pramanik}. For the above choice the optimal lower bound of Bob's 
uncertainty is lower bounded by\cite{Pramanik}
\begin{eqnarray}
\mathcal{L}_3(\rho_{AB}^{MM}) \approx 1.745.
\label{LB_3_MM}
\end{eqnarray}
When Bob chooses the classical strategy, for the above choice of observable 
his uncertainty given by (\ref{C_Strategy_QM}) is lower bounded by
\begin{eqnarray}
\mathcal{L}_0(\rho_{AB}^{MM})=2.
\label{LB_0_MM}
\end{eqnarray}
Employing the  quantum strategy, Bob's uncertainty (\ref{EUR-QM_Exp}) is lower bounded by 
\begin{eqnarray}
\mathcal{L}_1(\rho_{AB}^{MM}) \approx 1.5589,
\label{LB_1_MM}
\end{eqnarray}
whereas, the bound \cite{Pati} is given by
\begin{eqnarray}
\mathcal{L}_{2}(\rho_{AB}^{MM}) \approx 1.6226,
\label{LB_2_MM}
\end{eqnarray}
where the  classical information $C_B^M(\rho_{AB}^{MM})=0.1887$ (obtained from Eq.(\ref{Cinf_M_MM}) using our choice of $c_i$'s) and the quantum discord $D_B(\rho_{AB}^{MM})=0.2524$ (obtained from Eq.(\ref{D_MM})), tightens Berta's lower bound given by Eq.(\ref{LB_1_MM})\cite{QMemory}.

When both Alice and Bob measure same observable, $\sigma_z$ ($\sigma_x$) on their respective system, Bob extracts the information given by $C_{A,B}^{\sigma_z,\sigma_z}(\rho_{AB}^{MM})= 0.0659$ ($C_{A,B}^{\sigma_x,\sigma_x}(\rho_{AB}^{MM})= 0.1887$) for the above choice of $c_i$'s. Now, using Eq.(\ref{EUR-QM4}) Bob's uncertainty is 
lower bounded by
\begin{eqnarray}
\mathcal{L}_4(\rho_{AB}^{MM}) \approx 1.745,
\label{LB_4_MM}
\end{eqnarray}
which is equal to the optimal lower bound obtained using fine-grained 
uncertainty relation \cite{Pramanik}. One sees that though in this case, 
$\mathcal{L}_{2}(\rho_{AB}^{MM})$ tightens the bound $\mathcal{L}_1(\rho_{AB}^{MM})$,  it is not possible for either of them to be realized in practice since
they are not optimal. Fig.1 depicts the
main result of the paper, {\it viz.}, the optimal lower bound obtained
through the quantum strategy where the concept of extractable classical
information is applied. For the three classes of the states depicted, one
sees that the result
\begin{equation}
\mathcal{L}_1 \le \mathcal{L}_2 \le (\mathcal{L}_3 = \mathcal{L}_4)
\label{boundcompare}
\end{equation}
holds. 

\begin{figure}[!ht]
\resizebox{13cm}{8cm}
{\includegraphics{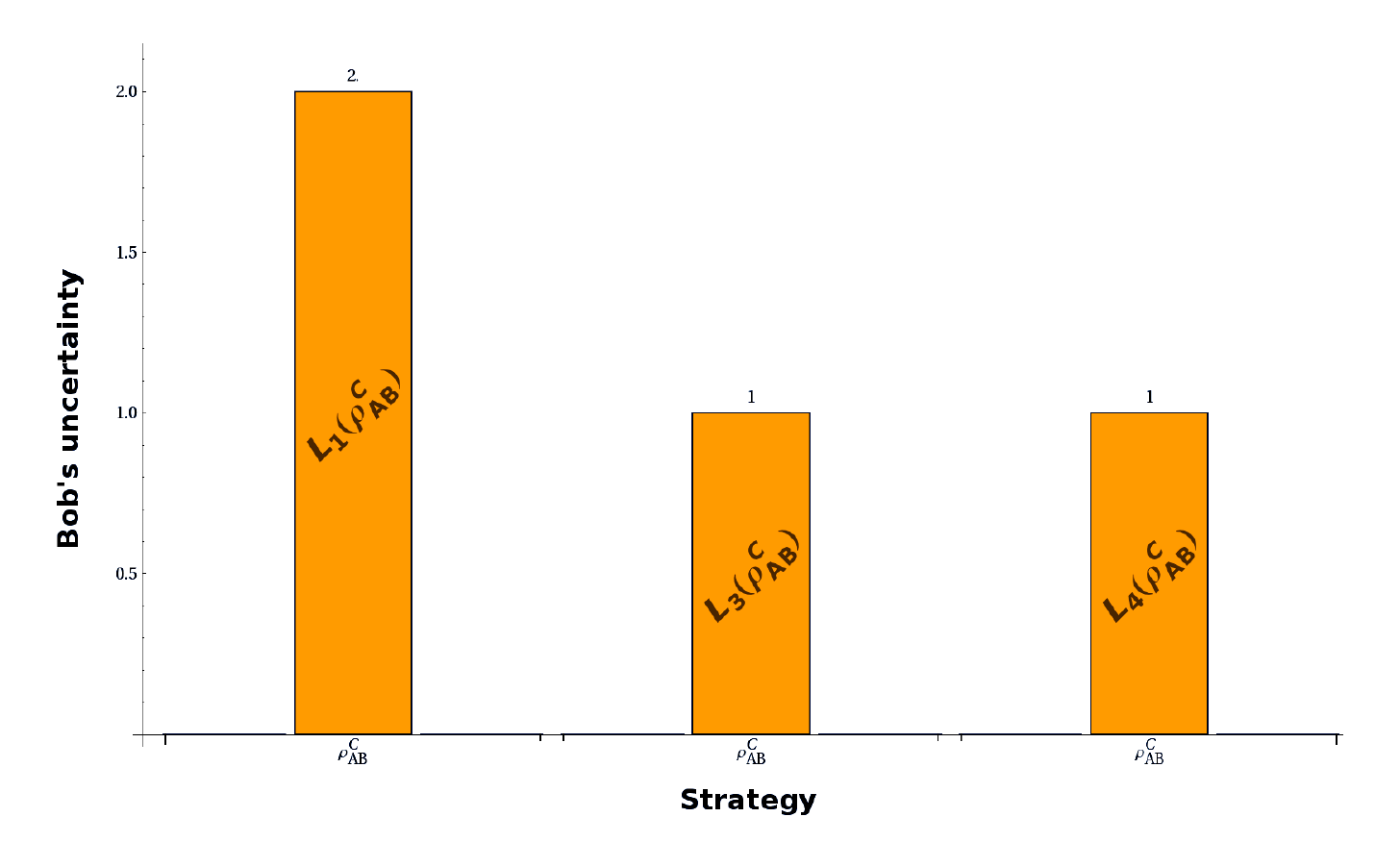}}
\caption{\footnotesize
A comparison of the  different lower bounds for the shared classical state 
chosing p=0.5.
}
\label{Fig.2}
\end{figure}

{\bf Classical state :} Now, we consider classical state $\rho_{AB}^C$, given by
\begin{eqnarray}
\rho_{AB}^C = p|00\rangle\langle 00| + (1-p) |11\rangle\langle 11|.
\end{eqnarray}
The state, $\rho_{AB}^C$ is a zero discord state \cite{ZDiscord}, i.e.,
\begin{eqnarray}
D_B(\rho_{AB}^{C})=0.
\end{eqnarray}
The classical information of the state $\rho_{AB}^C$ is given by
\begin{eqnarray}
C_B^M(\rho_{AB}^{C})=\mathcal{H}(p).
\end{eqnarray}
$C_B^M(\rho_{AB}^{C})$ gives the information about Alice's measurement outcome 
for the measurement of observable $\sigma_z$ to Bob, when Bob measures the 
same observable $\sigma_z$. 
The winning probability of the game characterized by winning condition 
$a\oplus b=1$ \cite{Pramanik} is
\begin{eqnarray}
P^{\text{game}}(\rho_{AB}^C)= \frac{\sin^2[\theta_S]}{2}.
\label{FUR_game_C}
\end{eqnarray}
Hence, the choices (for Alice) of the set of observables $\{R,S\}$ (which minimize Bob's uncertainty about Alice's outcome) are given by Eq.(\ref{Sett.}).

In this case, when Bob chooses the {\it classical strategy}, his uncertainty (given in Eq.(\ref{EUR-QM})) for the choices of settings given by Eq.(\ref{Sett.}) is lower bounded by an amount
\begin{eqnarray}
\mathcal{L}_0(\rho_{AB}^{C})=1+\mathcal{H}(p).
\label{LB_0_C}
\end{eqnarray}
When Bob applies the {\it quantum strategy} \cite{QMemory,Expt_M,Pati} his 
uncertainty is lower bounded by 
\begin{eqnarray}
\mathcal{L}_1(\rho_{AB}^{C})=\mathcal{L}_2(\rho_{AB}^{C})=1.
\label{LB_1_C}
\end{eqnarray}
For the state $\rho_{AB}^{C}$, Bob's extractable classical information 
(given by Eq.{\ref{CinfRR}) is $C_{A,B}^{\sigma_z,\sigma_z}(\rho_{AB}^{MM})= \mathcal{H}(p)$ ($C_{A,B}^{\sigma_x,\sigma_x}(\rho_{AB}^{MM})= 0$) when both of them measure 
the same observable $R=\sigma_z$ $(S=\sigma_x)$ on their respective particles. 
Hence, the lower bound given by Eq.(\ref{LB_4}) becomes
\begin{eqnarray}
\mathcal{L}_4{\rho_{AB}^{C}}= 1.
\end{eqnarray}
Finally, the optimal lower bound given by Eq.(\ref{LB_3}) is also
\begin{eqnarray}
\mathcal{L}_3{\rho_{AB}^{C}}= 1.
\end{eqnarray}
Hence in this case, $\mathcal{L}_1=\mathcal{L}_2=\mathcal{L}_3= \mathcal{L}_4=1 < \mathcal{L}_0$.
We thus observe that even purely classical correlations can play a role
in reducing the uncertainty using a shared bipartite state when the
quantum strategy is employed. This result is displayed in Fig.2. 

\section{Conclusions}

To summarize, in the present work we derive a new form of the uncertainty relation which
enables to reduce Bob's uncertainty maximally about Alice's measurement outcome while they choose same observable  with
the help of a shared state and communication between the two.  Using examples
of several classes of pure and mixed states of two qubits possessing 
quantum and classical correlations, we show that the lower 
bound of the uncertainty relation derived here is equal to the optimal lower 
bound (in the sense that Bob's uncertainty is reduced maximally) obtained with the help of the fine-grained uncertainty 
relation \cite{FUR,Pramanik}. We identify as the extractable classical 
information the
physical quantity  that
 is responsible for maximally reducing the uncertainty for the 
measurement of two non-commuting observables. Thus, the uncertainty relation
presented here provides an explanation in terms of physical resources as to
how the uncertainty of measurement of two incompatible observables may
be reduced maximally using shared correlations and classical communication.  
Our analysis further explains how the uncertainty may be reduced using the
quantum strategy even in the absence of quantum correlations when the two
parties share just a classically correlated state. 

Before concluding, it
will be worthwhile to stress that in the present work we have reformulated the 
uncertainty relation in the presence of quantum memory. The lower bound
of uncertainty is derived here using an approach that is different from the 
fine-graining employed earlier in the context of memory~\cite{Pramanik} and 
steering~\cite{Fur_Steer}. Though it turns out that for several important and 
widely considered examples of two-qubit states the bound derived here and 
that using fine-graining  turn out to be numerically equivalent, 
there is as yet no proof of formal equivalence between the two derived bounds using
two {\it a priori} different concepts of classical information and 
fine-graining, respectively. Further investigation into this issue is
called for in order to clarify  whether the connection
between fine-graining and extractable classical information (the validity of 
the last equality of our relation (\ref{boundcompare})) would hold true
for other classes of two-qubit states, or  could even be extended to the
case of higher dimensional systems.  Finally, it will be interesting to
extend our present analysis  in the light of other recent improvements in
the entropic uncertainty relations \cite{puchala2013}.

{\bf Acknowledgements:}\\
A.S.M. acknowledges support
from the project SR/S2/LOP-08/2013 of DST, India. T.P. acknowledges financial support from ANR retour des post-doctorants NLQCC (ANR- 12-PDOC-0022- 01).

\end{document}